\documentclass[reprint,amssymb,nobibnotes,showpacs,citeautoscript,longbibliography, floatfix, aps, prb]{revtex4-1}

\pdfoutput=1

\usepackage{graphicx}
\usepackage{xspace}

\usepackage{hyperref}
\hypersetup{colorlinks=true, linkcolor=blue, citecolor=blue, urlcolor=blue, unicode=false}

\newcommand{\BiSe}{Bi$_2$Se$_3$\xspace}
\newcommand{\BiTe}{Bi$_2$Te$_3$\xspace}

\begin{document}

\title{Direct observation of decoupled Dirac states at the interface between topological and normal insulators}

\author{M.~H.~Berntsen}
\altaffiliation{Present address: Deutsches Elektronen-Synchrotron (DESY), Notkestrasse 85, 22607 Hamburg, Germany}
\email{mhbe@kth.se}
\affiliation{KTH Royal Institute of Technology, ICT Materials Physics, Electrum 229, 164 40 Kista, Sweden}
\author{O.~G\"{o}tberg}
\affiliation{KTH Royal Institute of Technology, ICT Materials Physics, Electrum 229, 164 40 Kista, Sweden}
\author{B.~M.~Wojek}
\affiliation{KTH Royal Institute of Technology, ICT Materials Physics, Electrum 229, 164 40 Kista, Sweden}
\author{O.~Tjernberg}
\email{oscar@kth.se}
\affiliation{KTH Royal Institute of Technology, ICT Materials Physics, Electrum 229, 164 40 Kista, Sweden}

\date{\today}

\begin{abstract}
Several proposed applications and exotic effects in topological insulators rely on the presence of helical Dirac states 
at the interface between a topological and a normal insulator. In the present work, we have used low-energy 
angle-resolved photoelectron spectroscopy to uncover and characterize the interface states of \BiSe thin films and 
\BiTe/\BiSe heterostuctures grown on Si(111). The results establish that Dirac fermions are indeed present at the 
topological-normal-insulator boundary and absent at the topological-topological-insulator interface. Moreover, it is 
demonstrated that band bending present within the topological-insulator films leads to a substantial separation of the 
interface and surface states in energy. These results pave the way for further studies and the realization of 
interface-related phenomena in topological-insulator thin-film heterostructures.
\end{abstract}

\pacs{71.20.-b, 73.20.At, 73.40.Lq, 79.60.-i}
                
\maketitle

\section{Introduction}
Topological insulators (TIs) have attracted considerable attention in the recent years~\cite{Ando2013}. The 
three-dimensional (3D) TIs constitute a class of materials in which the inverted band structure of the insulating bulk 
is accompanied by the existence of metallic states at the surface~\cite{Fu2007a,Moore2007,Fu2007}. These surface states 
consist of massless helical Dirac fermions which are protected by time-reversal symmetry and which are therefore robust 
against non-magnetic backscattering. Although discovered only recently, TIs have been the subject of extensive research 
due to their inherently interesting properties and potential impact on the development of room-temperature spintronic 
devices~\cite{Yazyev2010,Steinberg2011}. Also, the realization of a number of exotic physical phenomena, such as 
Dyons~\cite{Qi2009}, Majorana fermions~\cite{Fu2008} and axion dynamics~\cite{Li2010d} is anticipated in these 
materials. 

While the TI surface states on \textit{in-situ} prepared clean or adsorbate-covered surfaces have been examined in 
detail (cf. Refs.~\onlinecite{Ando2013,Hasan2010} as well as the references therein), studies of the interactions of TIs 
with other materials, e.g. at heterostructure interfaces, are scarce up to now. Nonetheless, investigations of such 
interaction effects will no doubt lead to an even better understanding of the TIs and lay the foundation for the 
realization of concrete devices. In many semiconductor applications the properties of interfaces determine the operation 
characteristics of the device. Hence, studying the phenomena occurring at the technology-relevant TI-semiconductor 
interface and possible applications of TI thin films, rather than bulk crystals, is of particular interest.

Previous works on \BiSe thin films have revealed a strong variation in the electronic band structure of the surface 
state as a function of film thickness~\cite{He2010a}. When grown on a double-layer-graphene-terminated 6H-SiC(0001) 
substrate, at thicknesses below 5 quintuple layers (QL), one QL is 9.5~\AA{} thick, an energy gap opens at the Dirac 
point as a result of the hybridization between Dirac states originating from opposite sides of the film. The strength of 
the hybridization decreases rapidly with increasing film thickness until it is negligible for 6~QL films and the energy 
gap at the Dirac point closes. The experimental observation of a gap in ultra-thin films provides indirect evidence for 
the existence of a Dirac state at the interface towards the substrate. Similar observations have been made in 
(PbSe)$_5$(\BiSe)$_{3m}$ single crystals, where natural heterostructures consisting of TI (\BiSe) and normal-insulator 
(PbSe) layers are formed~\cite{Nakayama2012}. In fact, theory predicts the existence 
of gapless topological Dirac states at interfaces between topologically trivial materials and TIs~\cite{Song2010}. 
Consequently, such states should exist at the interface between a TI thin film and an insulating substrate on which it 
is grown. However, so far no direct observation of buried \emph{non-hybridized} interface states in TI films has been 
made. Therefore, it is of fundamental interest to experimentally establish the existence of interface states in a 
TI-semiconductor junction by more direct means. Additionally, a qualitative description of how the electron 
configuration in thin films is influenced by the presence of the substrate, including a band-alignment model for the 
junction, is desirable, something which thus far has only been discussed briefly~\cite{He2010a,Sakamoto2010,Shan2010}. 
The latter can provide valuable insights on how to ``tailor'' the electronic properties of a TI film, e.g. the position 
of the chemical potential, by changing properties of the substrate. Such an understanding is 
the foundation needed to uncover novel phenomena as well as to realize practical device applications. Here we address 
these issues through photoemission studies of TI thin films and TI thin-film heterostructures grown on silicon 
substrates, provide experimental evidence for the existence of decoupled Dirac fermions at the substrate interface and 
show how the band bending across the film can be explained by band alignment between the film and the substrate at the 
interface. 

\section{Experimental details}
In the present work we have studied the electronic band structure of \BiSe thin films and \BiTe/\BiSe thin-film 
heterostructures by means of laser-based angle-resolved photoelectron spectroscopy (ARPES). The experiments were 
conducted at the BALTAZAR laser-ARPES facility~\cite{Berntsen2011} (KTH, Stockholm, Sweden) using an angle-resolving 
time-of-flight electron analyzer and a photon energy of 10.5~eV. The data were collected at $T = 9$~K and at a base 
pressure of $5\times10^{-11}$~mbar. The combination of the low excitation energy and the cryogenic sample temperature 
increases the escape depth of the photoelectrons~\cite{Seah1979}, thus making the direct observation of buried interface 
states feasible.

Thin films of (0001)-oriented \BiSe were grown \textit{in situ} by co-evaporation of Bi and Se onto a Bi-terminated 
Si(111)-($7\times 7$) substrate following the method presented in Ref.~\onlinecite{Zhang2009a} which produces atomically 
flat, high quality, stoichiometric films. Two different types of Si(111) substrates were used, one arsenic doped 
($n$-type) with a resistivity of 4~m$\Omega\cdot$cm (determined by a Hall measurement), and one boron doped ($p$-type) 
with a resistivity of 0.9~m$\Omega\cdot$cm (Si-Mat Silicon Materials, Germany). The substrates were prepared by repeated 
cycles of annealing at 1100~$^{\circ}$C in order to achieve the ($7\times 7$) surface reconstruction. The subsequent 
deposition of one monolayer (1~ML) Bi at a substrate temperature of 500~$^{\circ}$C formed a Si(111)$\beta \sqrt{3} 
\times \sqrt{3}$-Bi surface on which the \BiSe film could be grown. The \BiSe thin films were grown by co-evaporation of 
high purity Bi (99.999~\%) and Se (99.999~\%) (both from Goodfellow Cambridge Ltd., United Kingdom) using an 
electron-beam evaporator while keeping the substrate temperature at 270~$^{\circ}$C. A quartz-crystal monitor was used 
to set the growth rates of Bi and Se prior to the deposition and the built-in flux monitor of the evaporator ensured a 
constant rate during the deposition. Repeated tests showed that the best film quality was achieved when using a large Se 
overdose. The typical growth rate and Se overdose for the films in this study were about 2.5~\AA{} per minute and 80~\%, 
respectively. The procedure used for the Bi-termination of the Si(111)-($7\times 7$) surface was identical for both 
types of substrates and all \BiSe films were deposited under similar conditions. 

Thin-film \BiTe/\BiSe heterostructures were manufactured by deposition of 2~QL \BiTe on top of pre-grown 6~QL \BiSe 
films. The procedure for growing \BiTe films was similar to the \BiSe case, i.e. with similar substrate temperature and 
a deposition rate of Te comparable to that of Se in the previous cases. During the preparation of the substrates and 
growth of the films the base pressure in the vacuum chamber was better than $7\times10^{-10}$~mbar. 

\section{Results}

\begin{figure*}
\includegraphics[width=.9\textwidth]{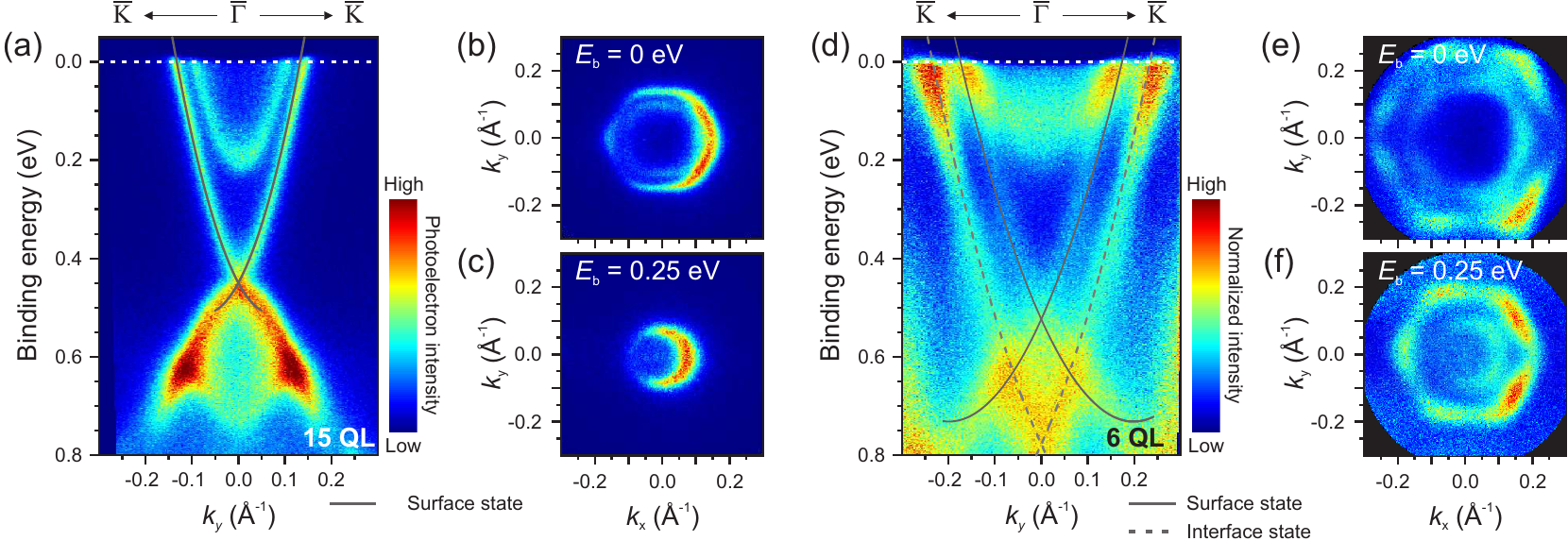}
\caption{(Color online) Photoemission intensity plots. (a)--(c) Raw photoemission data of a $\sim$15~QL thick \BiSe 
film on an $n$-type Si-substrate. The data were taken at $h\nu = 10.5$~eV and $T = 9$~K. (a) Energy dispersion along 
the $\bar{\text{K}}$-$\bar{\Gamma}$-$\bar{\text{K}}$ direction. The fitted band structure described by 
Eq.~(\ref{singleband}) is plotted as solid lines on top of the experimental data. (b) Constant-energy cut at 
$E_{\text{F}}$ displaying the hexagonal Fermi surface. (c) Constant-energy cut at $E_{\text{b}}=0.25$~eV. (d)--(f) 
Normalized photoemission data of a 6~QL \BiSe film on an $n$-type Si-substrate taken at $h\nu = 10.5$~eV and $T = 9$~K. 
All energy distribution curves (EDCs) in the data set are normalized to unity in order to increase the visibility of 
the states. (d) Measured energy dispersion along the $\bar{\text{K}}$-$\bar{\Gamma}$-$\bar{\text{K}}$ direction 
together with the fitted band structure of Eq.~(\ref{doubleband}). The solid and dashed lines represent the surface and 
interface states, respectively. (e) Constant-energy cut at $E_{\text{F}}$. Two hexagonal energy contours are observed, 
the inner one belonging to the surface state and the outer to the interface state. (f) Constant-energy cut at 
$E_{\text{b}} = 0.25$~eV. The black areas in (e) and (f) are outside the detector.}
\label{arpes}
\end{figure*}

We first turn our attention to the \BiSe films. Figure~\ref{arpes} presents photoemission data of two films with 
thicknesses $\sim$15~QL [panels (a)--(c)] and 6~QL [panels (d)--(f)] grown on $n$-type Si(111) substrates. Both films 
are sufficiently thick so that hybridization between Dirac states at opposite surfaces of the films is avoided. While 
the 6~QL film has a thickness which could allow the topologically protected state at the interface towards the substrate 
to be directly probed by a low-photon-energy ARPES experiment, the $\sim$15~QL film is too thick for the interface state 
to be directly observed. Therefore, the ``thick'' film serves as a reference for the dispersion of the surface state. 
The energy-momentum slice displayed in Fig.~\ref{arpes}(a) shows a nearly linearly dispersing Dirac surface state with 
the Dirac point located at a binding energy $E_{\text{b}} = 0.45$~eV. The parabolic band at lower binding energy is a 
quantum-well state resulting from the confinement of conduction-band electrons due to the finite thickness of the film. 
As revealed by the constant-energy cut taken at the Fermi level ($E_{\text{F}}$), shown in Fig.~\ref{arpes}(b), the 
Fermi surface is hexagonally deformed. Closer to the Dirac point the constant-energy surface becomes circular, as seen 
in Fig.~\ref{arpes}(c). This deviation from an ideal Dirac cone has been reported previously~\cite{Kuroda2010} and is 
expected for highly $n$-doped \BiSe samples. Following Ref.~\onlinecite{Lu2010} the energy dispersion of the surface 
state in the thick film can be described by a single-band model:
\begin{equation}
 E_{\pm}(k)=E_0-Dk^2\pm\sqrt{\left(\frac{\Delta}{2}-Bk^2\right)^2+\left(v_{\text{D}} \hbar k\right)^2}.
 \label{singleband}
\end{equation}
Here, $E_{\pm}$ represent the dispersion of the Dirac state above and below the Dirac point, respectively, $E_0$ is the 
binding energy of the Dirac point, $\Delta$ is the energy gap at the Dirac point (non-zero only if inter-surface 
coupling is present), $v_{\text{D}}$ is the band velocity in the vicinity of $\bar{\Gamma}$ and $\hbar k$ is the 
in-plane crystal momentum. $D$ and $B$ are coefficients of quadratic terms of different origins. While the $D$ term is a 
result of the broken particle-hole symmetry in the bulk valence and conduction bands causing a parabolic correction to 
the surface state~\cite{Zhang2012}, the $B$ term characterizes massive states in the presence of a 
gap~\cite{Zhang2009b}. The latter term always vanishes in our considerations of non-hybridized topologically protected 
states.

Fitting the model to extracted momentum-distribution-curve (MDC) peak positions results in the solid lines plotted on 
top of the photoemission spectrum presented in Fig.~\ref{arpes}(a). The corresponding values of the model parameters are 
listed in Tab.~\ref{table1}. The model described by Eq.~(\ref{singleband}) is derived from an effective Hamiltonian and 
fits well with the data for small $k$ values. Any model describing the surface-state dispersion more accurately would 
have to include further higher-order terms. In particular, a $k^3$ term is needed to account for the hexagonal 
warping~\cite{Fu2009}. To keep the model as simple as possible, we neglect these effects here and concentrate on the 
region of small $k$, only. The hexagonal distortion of the Dirac cone in this case is visible from approximately 250~meV 
above the Dirac point up to $E_{\text{F}}$. Therefore, only points from the region where no distortion of the cone is 
observed ($200~\mathrm{meV} \leqslant E_{\mathrm{b}} \leqslant 500~\mathrm{meV}$) are included in the fit. We note, that 
the resulting Dirac velocity ($v_{\text{D}}$) is somewhat smaller than reported elsewhere~\cite{He2010a,Sakamoto2010}. 
For comparison, we quote both $v_{\text{D}}$ as well as the Fermi velocity ($v_{\text{F}}$) in Tab.~\ref{table1}.

The comparison of the energy-momentum slices in Figs.~\ref{arpes}(a) and \ref{arpes}(d) for the thick and the 6~QL 
films, respectively, reveals the presence of similar Dirac-like states in both films with the Dirac points located at 
nearly the same binding energies. However, in the 6~QL data there is an additional, outer, V-shaped feature reminiscent 
of a Dirac state shifted towards higher $E_{\text{b}}$. Note that the data displayed in Figs.~\ref{arpes}(d)--(f) are 
normalized in order to improve the visibility of the features in the spectrum [all energy distribution curves (EDCs) are 
normalized to yield the same integrated intensity]. The constant-energy slice displayed in Fig.~\ref{arpes}(e) is taken 
at $E_{\text{F}}$ and shows that both features are hexagonally deformed similar to what we observe in the thick film. 
For a constant-energy cut at higher binding energy the inner part becomes circular, as seen in Fig.~\ref{arpes}(f), 
since we are now approaching the Dirac point of this state, while the outer 
constant-energy surface remains hexagonal. In addition to the Dirac states a parabolic conduction-band quantum-well 
state located at lower binding energy is observed. The model fitted to the thick-film data describes a single Dirac 
state only. Therefore, in order to investigate whether or not the additional spectral feature seen in the 6~QL case is 
in fact the interface state, we adopt the double-band model from Ref.~\onlinecite{Shan2010}. This model includes Dirac 
cones located on opposite sides of the TI film and accounts for the structure inversion asymmetry (SIA) and the 
effective band bending induced by the presence of the substrate. The dispersion of the Dirac states is then given by
\begin{equation}
 E_{\sigma\pm}(k)=E_0-Dk^2\pm\sqrt{\left(\frac{\Delta}{2}-Bk^2\right)^2+\left(\vert V\vert+\sigma v_{\text{D}} \hbar 
k\right)^2}.
 \label{doubleband}
\end{equation}
The term $\vert V\vert$ represents the band bending across the film and together with $\sigma = \pm 1$ the model 
produces a set of two Dirac cones with opposite helicities having their Dirac points separated in energy by $2\vert 
V\vert$. $E_0$ is now the center energy of the two Dirac points. The remaining parameters have the same meaning as in 
Eq.~(\ref{singleband}).

Fitting Eq.~(\ref{doubleband}) to the data of the 6~QL film reproduces the observed band structure well, as seen in 
Fig.~\ref{arpes}(d). Also in this case, only data points close to the Dirac points are used in the fit. The outer 
V-shaped state observed in the simulated band structure [dashed line in Fig.~\ref{arpes}(d)] corresponds to the Dirac 
cone located at the interface between the TI and the substrate. This suggests that we directly observe the interface 
state from the ``bottom'' surface of the TI film. Also, the absence of an outer Dirac-like state in the thick film is 
consistent with our interpretation regarding the interface state since this film is too thick for the electronic state 
at the interface to be directly probed in the experiment.

\begin{figure}
\includegraphics[width=\columnwidth]{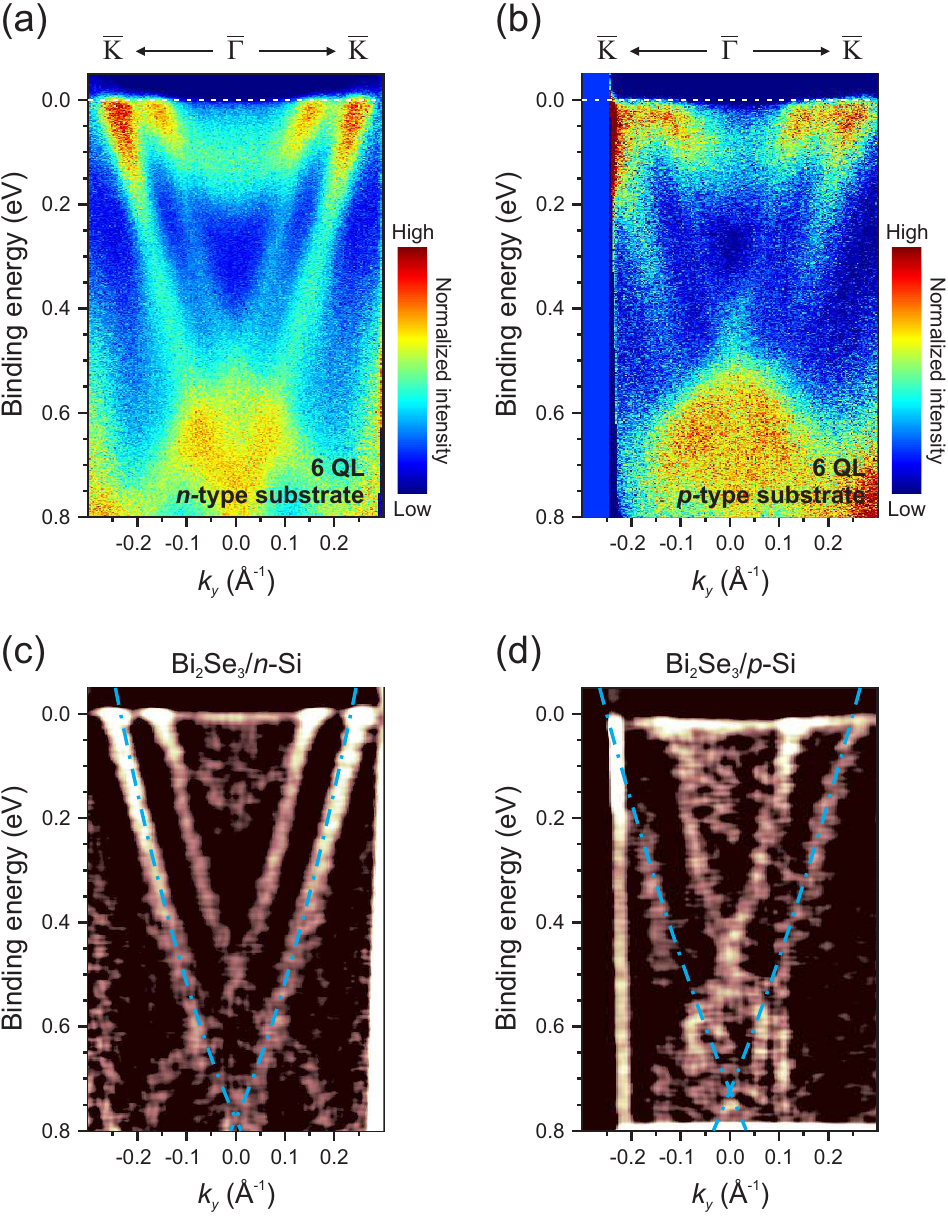}
\caption{(Color online) \BiSe thin films grown on differently doped Si substrates. (a) Normalized intensity plot of the 
energy dispersion along the $\bar{\text{K}}$-$\bar{\Gamma}$-$\bar{\text{K}}$ direction for 6~QL \BiSe grown on an 
$n$-type substrate. (b) Corresponding energy-momentum slice for a 6~QL film on a $p$-type substrate (also normalized 
intensity). Due to missing data for $k_{y}$ values below $-0.25~\text{\AA}^{-1}$ the intensity in this region has been 
manually set to zero. (c)--(d) Second-derivative plots corresponding to the spectra shown in (a) and (b), respectively. 
Non-normalized data are used as input and the second derivative is calculated by convolution with a Laplacian kernel. 
The dash-dotted lines represent the interface state calculated using Eq.~(\ref{doubleband}) and the parameters listed in 
Tab.~\ref{table1}.}
\label{comparison}
\end{figure}

The data of the 6~QL film reveals that the electronic state located at the interface towards the substrate exhibits a 
similar Dirac-like dispersion as the surface state. The binding-energy shift of the interface state, with respect to the 
surface state, is indicative of a band bending in the TI film, likely to be caused by the presence of the substrate. 
This observation motivates the further investigation of the ``substrate effect'', i.e. the influence of the choice of 
the substrate on the band bending and electronic configuration of the TI film. Already at this stage, our results 
suggest that the band bending for a 6~QL \BiSe film grown on an $n$-Si(111) substrate is larger than for films with the 
same thickness grown on double-layer-graphene-terminated 6H SiC(0001) substrates~\cite{He2010a} (270~meV compared to 
136~meV, respectively). To gain a qualitative understanding of the substrate-induced effect, we continue by comparing 
6~QL films grown on $n$-type (4~m$\Omega\cdot$cm) and $p$-type (0.9~m$\Omega\cdot$cm) Si(111) substrates. Interestingly, 
as seen in Fig.~\ref{comparison}, this contrasting juxtaposition reveals very little differences between the observed 
band structure of the films although the bulk positions of $E_{\text{F}}$ in the two substrates are very dissimilar. In 
both cases the energy separation between the two Dirac points ($\Delta E_{\text{D}}$) is roughly 0.3~eV. The striking 
similarities between the 6~QL films on $n$-type and $p$-type substrates are consequences of the Fermi level of the 
Si(111)-($7\times 7$) surface being pinned approximately 0.7~eV above the valence-band maximum ($E_{\text{V}}$) due to a 
high density of states located in the band gap at the surface~\cite{Himpsel1981,Himpsel1983,Nicholls1987}. This pinning 
of $E_{\text{F}}$ is independent of the bulk doping and thus equivalent for both $n$-type and $p$-type substrates. 
Terminating the surface with 1~ML bismuth has little effect on the pinning level, reducing $E_{\text{F}}-E_{\text{V}}$ 
from 0.7~eV to approximately 0.65~eV (cf. Ref.~\onlinecite{Hricovini1991}).

While the overall band dispersion seen in the ARPES spectra in Fig.~\ref{comparison} essentially does not change for the 
films grown on different substrates, one still notices pronounced intensity differences in the spectra. We ascribe these 
to slight variations in the quality of the film boundaries. It is known that moderate surface disorder leads to a loss 
of spectral weight in the metallic surface states. This has been shown for the \BiSe system both experimentally using 
ion-bombardment-induced surface defects~\cite{Hatch2011} and theoretically by numerical simulations~\cite{Schubert2012}. 
In this respect, some degree of variation is expectable across different samples. This is true in particular for the 
visibility of the buried-interface state, since not always a perfect epitaxial growth can be guaranteed right from the 
start of the film deposition.

\begin{table}
\begin{tabular*}{\columnwidth}{@{\extracolsep{\fill}}  lccc}
\multicolumn{4}{c}{}   \\ \hline \hline
Film thickness & 6 QL\footnotemark[1] & 6 QL\footnotemark[1]  & $\sim$15 QL\footnotemark[2] \\ \hline
Substrate doping & $n$-type & $p$-type & $n$-type \\
$E_0$ (eV) &-0.661 &-0.580 &-0.450  \\ 
$D$ (eV\,\AA$^2$) &-5.84 &-2.23 & -12.4 \\
$B$ (eV\,\AA$^2$) &0 &0 &0 \\
$\Delta$ (eV)  &0 &0 &0 \\
$\vert V\vert$ (eV)& 0.135&0.144 &- \\
$\hbar{}v_{\text{D}}$ (eV\,\AA{}) &2.16 &2.33 & 1.75 \\
$v_{\text{D}}$ (10$^5$ m\,s$^{-1}$) &3.28 &3.54 & 2.65 \\
$\hbar{}v_{\text{F}}$ (eV\,\AA{})\footnotemark[3] &3.8 &3.2 &3.9 \\
$v_{\text{F}}$ (10$^5$ m\,s$^{-1}$)\footnotemark[3] &5.8 &4.8 &5.9 \\\hline \hline
\end{tabular*}
\footnotetext[1]{Fitted using Eq.~(\ref{doubleband}).}
\footnotetext[2]{Fitted using Eq.~(\ref{singleband}).}
\footnotetext[3]{Value along the $\bar{\Gamma}$-$\bar{\text{K}}$ direction, based on MDC peaks close to $E_{\text{F}}$.}
\caption{Fitted model parameters for Eqs.~(\ref{singleband}) and~(\ref{doubleband}).}
\label{table1}
\end{table}

\begin{figure}
\includegraphics[width=\columnwidth]{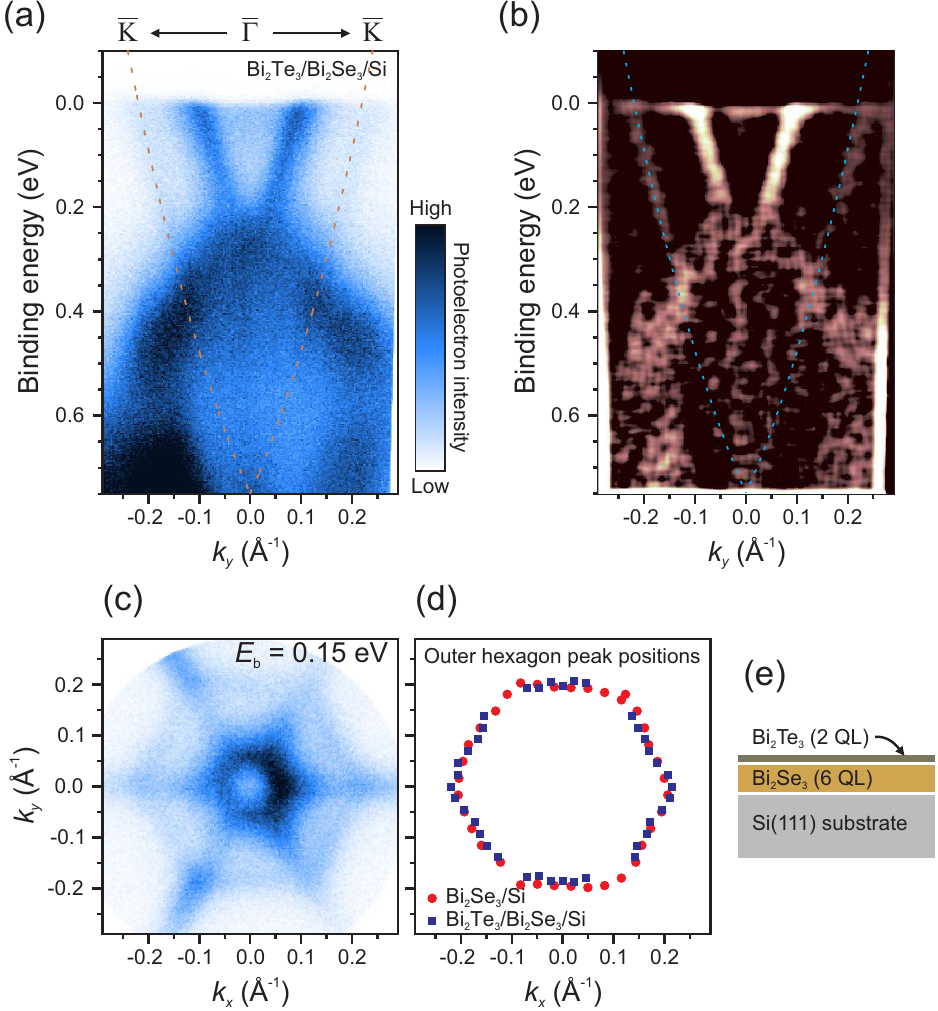}
\caption{(Color online) \BiTe/\BiSe/Si heterostructure. (a) Energy dispersion along the 
$\bar{\text{K}}$-$\bar{\Gamma}$-$\bar{\text{K}}$ direction for 2~QL \BiTe grown on a 6~QL \BiSe film. The thin, dotted 
line shows the interface state described by Eq.~(\ref{doubleband}) using the model-parameter values extracted from the 
6~QL \BiSe film data displayed in Fig.~\ref{arpes}(d). (b) Second-derivative plot of the energy dispersion along 
$\bar{\text{K}}$-$\bar{\Gamma}$-$\bar{\text{K}}$. (c) Constant-energy surface for a binding energy of 0.15~eV. The outer 
hexagon is characteristic of the state at the interface between the Si substrate and the \BiSe film, the inner hexagon 
represents the surface state mainly stemming from the \BiTe top layer. (d) Extracted constant-energy contour of the 
outer hexagon displayed in (c) (squares) compared to the corresponding contour from the data set of the 6~QL \BiSe film 
presented in Fig.~\ref{arpes} (circles). (e) Schematic drawing of the TI thin-film heterostructure.}
\label{heterostructure}
\end{figure}

In the next step, we turn to the study of a \BiTe/\BiSe thin-film heterostructure. By depositing a thin layer (2~QL) of 
\BiTe on top of a 6~QL \BiSe film we were able to study a TI film with \emph{a priori} distinct vacuum- and 
substrate-interface electronic properties. While a free-standing 2~QL \BiTe film is too thin to exhibit metallic surface 
states~\cite{Liu2012}, within the TI-TI heterojunction both layers can be regarded as topologically non-trivial without 
any Dirac state forming at the interface between them. The corresponding ARPES data are displayed in 
Fig.~\ref{heterostructure}. The energy-momentum spectrum, presented in Figs.~\ref{heterostructure}(a) 
and~\ref{heterostructure}(b), now features a surface state which has a distinct \BiTe character where the Dirac point at 
$\bar{\Gamma}$ lies below the valence-band maximum along the $\bar{\Gamma}$-$\bar{\text{K}}$ direction (cf. 
Ref.~\onlinecite{Zhou2012} for a direct comparison of the band structure of \BiSe and \BiTe). The additional ``outer'' 
feature of the 6~QL \BiSe film, cf. Fig.~\ref{arpes}(d), remains visible, however, with substantially less spectral 
weight as compared to the surface state. The reduced intensity of this feature is expected for the signal from a buried 
interface state between the Si substrate and the \BiSe film. Since the heterostructure film has a total thickness of 
8~QL the relative intensity of the interface state to the surface state is expected to be smaller as compared to the 
6~QL \BiSe film. This is consistent with our observations.

Similar to the 6~QL film a constant-energy surface from the data of the heterostructure sample displays a hexagonally 
shaped contour of the outer state, cf. Fig.~\ref{arpes}(e) and Fig.~\ref{heterostructure}(c), respectively. The inner 
feature in Fig.~\ref{heterostructure}(c) is star-shaped with ``arms'' along the $\bar{\Gamma}$-$\bar{\text{M}}$ 
directions, characteristic for \BiTe in the energy range close to the Dirac point~\cite{Chen2009}. Using the parameters 
listed in Tab.~\ref{table1} for the 6~QL \BiSe film on the $n$-type substrate and Eq.~(\ref{doubleband}), we only plot 
the band assigned to the interface state on top of the data in Fig.~\ref{heterostructure}(a). When slightly shifting the 
energy scale of the calculated band by 40~meV towards lower binding energies the model and data are in overall good 
agreement [see also the second-derivative plot in Fig.~\ref{heterostructure}(b)]. This small energy shift in the 
interface state between the two films can very well be related to minor film-to-film variations in the pinning position 
of $E_{\text{F}}$ at the substrate interface. Furthermore, the contour of the large hexagon 
in Fig.~\ref{heterostructure}(c) agrees well with the one from a corresponding constant-energy surface of the 6~QL 
\BiSe film, see Fig.~\ref{heterostructure}(d). In both cases the constant-energy slice is taken at the same energy above 
the Dirac point. 

\begin{figure*}
\includegraphics[width=.9\textwidth]{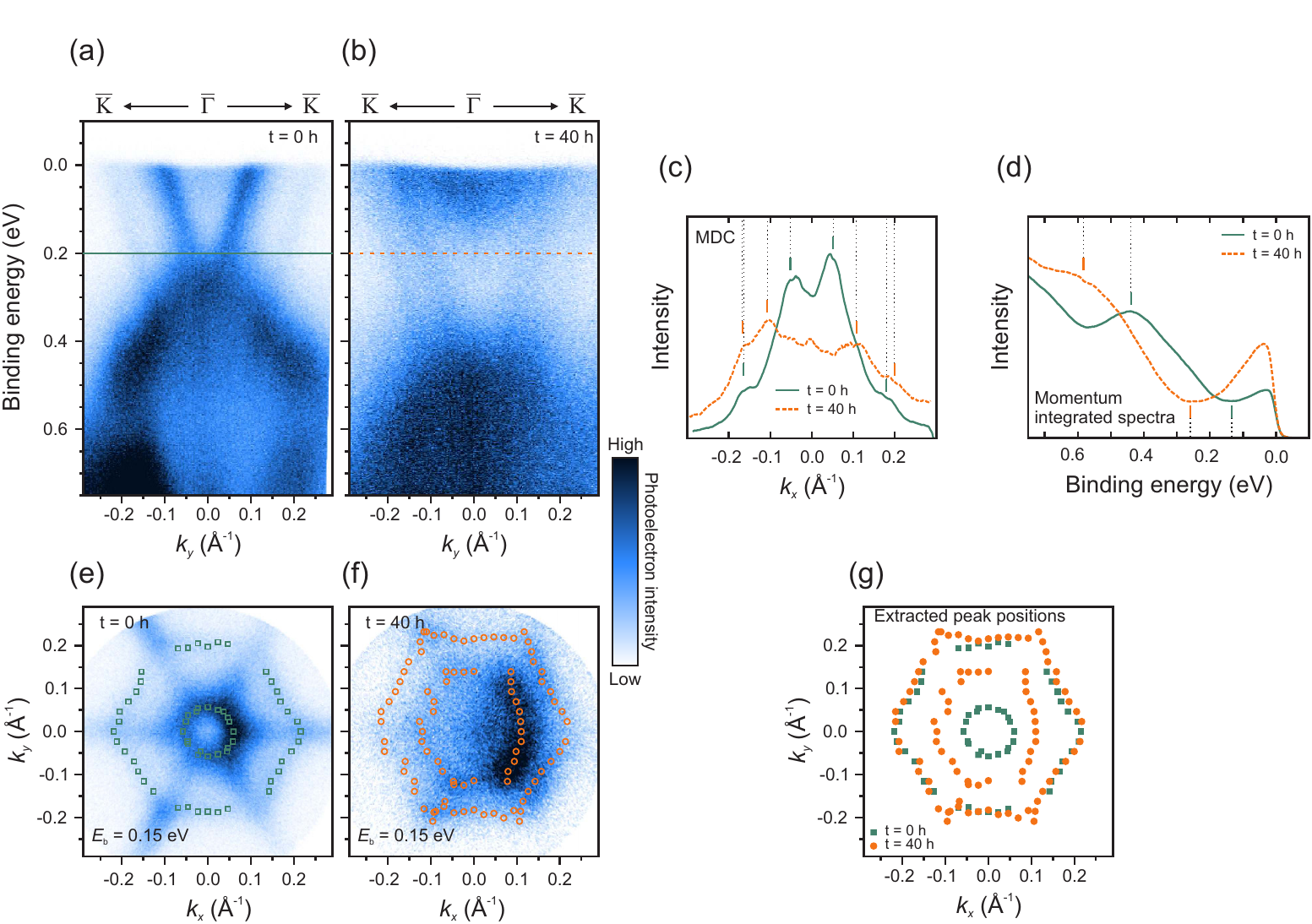}
\caption{(Color online) Time evolution of the surface state. (a) Energy-momentum dispersion along the indicated 
high-symmetry direction for the \BiTe/\BiSe/Si heterostructure directly after sample preparation ($t = 0$~h). (b) 
Corresponding energy-momentum plot for the evolved state at $t = 40$~h. (c) MDCs for $t = 0$~h (solid line) and $t = 
40$~h (dashed line) taken at the binding energy position indicated in (a) and (b), respectively. The indicated markings 
correspond to the center positions of fitted Lorentzians. (d) Intensity integrated over all momenta as a function of 
energy. The spectra are normalized to the value of the local minimum below the Fermi level. (e)--(f) Peak positions of 
the inner and outer states extracted from constant-energy slices of the measurements at $t = 0$~h and $t = 40$~h, 
respectively. (g) Comparison of the contours from (e) and (f).}
\label{evolution}
\end{figure*}

The clear similarities between the outer states in the two films support the notion of an interface state with \BiSe 
character. To further test this conclusion, we investigated the effects of sample aging on the two distinct electronic 
states in the heterostructure. For single crystals of \BiSe or \BiTe, cleaved in vacuum, adsorption of residual gases on 
the sample surface results in a band bending close to the surface which shifts the Dirac point of the surface state 
towards higher binding energies. This effect is time-dependent and depends on the rate of adsorption. The surface-state 
evolution has been studied by means of natural aging~\cite{King2011}, i.e. leaving the sample in vacuum for a period of 
time, as well as by controlled dosing of CO, H$_2$ and H$_2$O~\cite{Zhou2012, Benia2011}. As seen in 
Fig.~\ref{evolution}, in the \BiTe/\BiSe/Si heterostructure we observe a similar effect on the \BiTe-like state when the 
sample is kept in vacuum (base pressure $8\times10^{-11}$~mbar, $T=9$~K) for a period of 40~hours, whereas the 
\BiSe-like state is static, i.e. it shows no temporal evolution. 

Figure~\ref{evolution}(a) displays the energy dispersion along the $\bar{\text{K}}$-$\bar{\Gamma}$-$\bar{\text{K}}$ 
direction of the thin-film heterostructure directly after the sample preparation ($t = 0$~h). The corresponding spectrum 
for the evolved state ($t = 40$~h) is shown in Fig.~\ref{evolution}(b). Directly after the preparation the surface state 
has a clear \BiTe character and the Dirac point is located approximately at a binding energy of 250~meV to 300~meV. An 
MDC at $E_{\text{b}} = 200$~meV [solid line in Fig.~\ref{evolution}(c)] reveals the existence of the interface state as 
two low-intensity satellite peaks at approximately $\pm 0.17~\text{\AA{}}^{-1}$, flanking the dominant surface-state 
peaks at $\pm 0.05~\text{\AA{}}^{-1}$. An MDC from the evolved spectrum at the same binding energy [dashed line in 
Fig.~\ref{evolution}(c)] indicates that the main peaks have shifted towards higher $k$ values, implying that the Dirac 
point has moved down in energy (towards higher $E_ {\text{b}}$). Moreover, from the momentum-integrated spectra in 
Fig.~\ref{evolution}(d) it becomes apparent that the intense ``bump'' close to $E_{\text{b}} = 450$~meV at $t = 0$~h has 
moved to $E_{\text{b}} = 600$~meV in the $t = 40$~h measurement. The local minimum, located at lower binding energy, 
experiences a similar shift. Besides, additional spectral weight appears close to $E_{\text{F}}$. This is attributed to 
quantized conduction-band states which become occupied due to the shift of the chemical potential. The added intensity 
close to $E_{\text{F}}$ is also visible in Fig.~\ref{evolution}(b). 

By extracting the peak positions from the constant-energy surfaces at $E_{\text{b}} = 150$~meV of both the initial and 
the evolved data sets the contours of the outer and inner states can be identified, see Figs.~\ref{evolution}(e) 
and~\ref{evolution}(f), respectively. The direct comparison of these contours, presented in Fig.~\ref{evolution}(g), 
shows that the outer state is stationary while the circumference of the inner contour increases with time, consistent 
with the Dirac point moving towards higher binding energies. The fact that one of the states (the one carrying the 
characteristics of the \BiTe top layer exposed to the vacuum) shows a time evolution, while the other one (being very 
similar to the outer state in the 6~QL \BiSe film) is static, thus provides further strong evidence that the latter 
state is located at the \BiSe/substrate interface where the residual-gas adsorption has negligible effects. Eventually, 
we note that we do not observe any particular states arising from the presence of the \BiTe/\BiSe interface.

\begin{figure}
\includegraphics[width=.9\columnwidth]{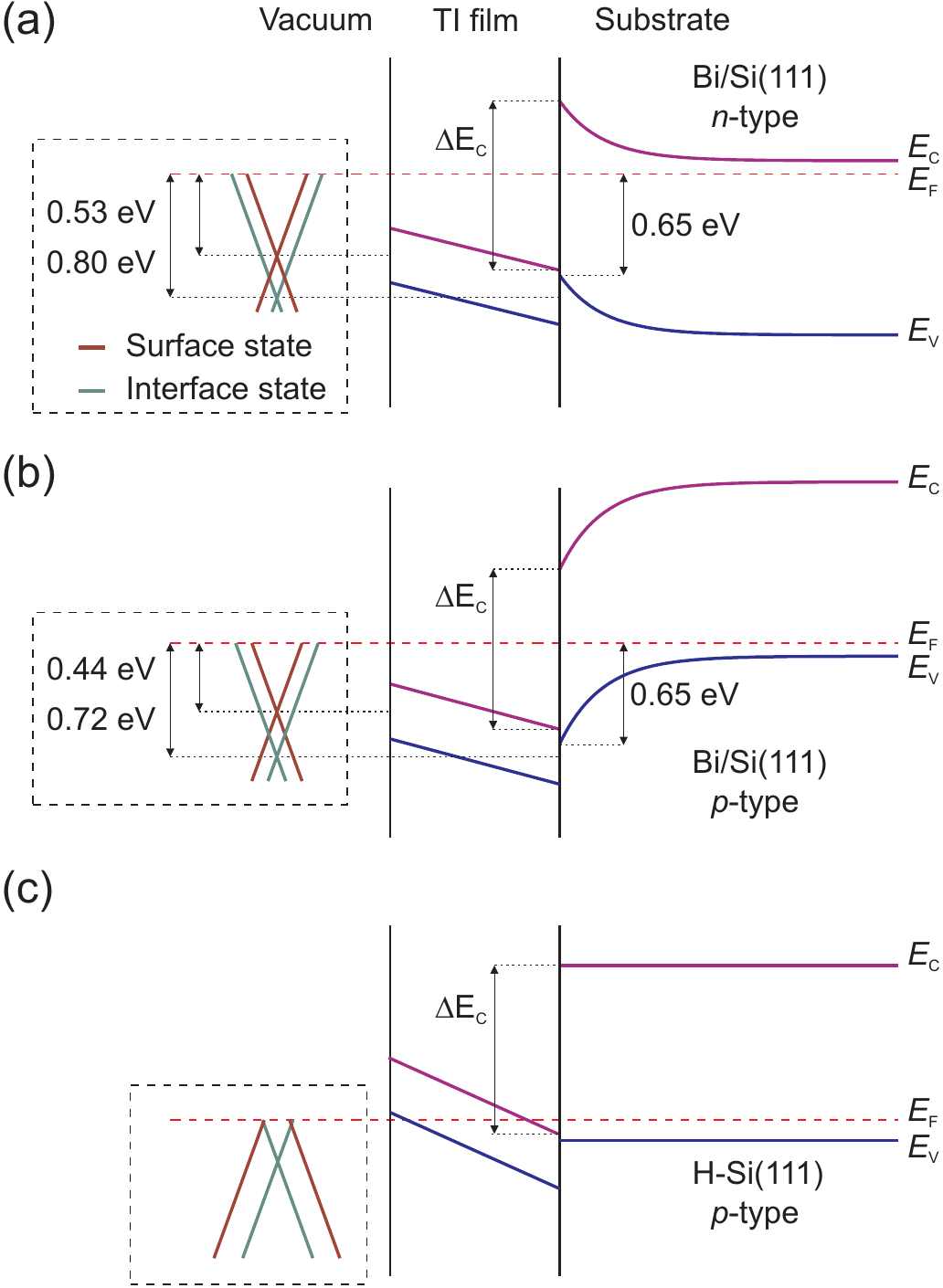}
\caption{(Color online) Relative energies of bulk bands in the TI films and substrates. (a) Band diagram for the 
\BiSe/Bi/$n$-Si system showing the conduction ($E_{\text{C}}$) and valence ($E_{\text{V}}$) bands in the thin film and 
substrate. The conduction-band offset $\Delta E_{\text{C}}$ at the interface is 1.09~eV. The dashed horizontal line 
refers to the Fermi level, $E_{\text{F}}$. Vertical lines represent the vacuum-TI and TI-substrate interfaces, 
respectively. A schematic drawing of the observed band structure for the surface and interface states with corresponding 
values for $E_{\text{D}}$ is enclosed by the dashed rectangle. (b) Band diagram for the \BiSe/Bi/$p$-Si system for which 
$\Delta E_{\text{C}}=1.02$~eV. (c) Predicted band diagram for a $p$-doped \BiSe thin film on a H-terminated Si(111) 
substrate.}
\label{heterjunct}
\end{figure}
\section{Discussion and conclusions}
Altogether, the agreement between the theoretical model and the experimental observation of the states in the 6~QL films 
combined with the unchanged nature of the outer state after deposition of another TI on the surface of the film support 
the notion that we, in our ARPES experiments, directly observe the Dirac state located at the substrate interface. The 
fact that no time-dependent energy shift is observed for the outer state in the heterostructure sample further supports 
its interface assignment. Consequently, returning to the \BiSe films in Fig.~\ref{comparison}, the extracted values of 
the energy difference between the Dirac points ($\Delta E_{\text{D}}$) of the surface and interface states, 
respectively, directly reveal the magnitude of the band bending across the films. This knowledge enables us to construct 
a band diagram for the TI-semiconductor junction.

By using the binding-energy positions of the Dirac points, together with the known band gap of \BiSe (0.35~eV, cf. 
Refs.~\onlinecite{Black1957,Xia2009}), the Fermi level relative to the conduction-band minimum can be determined. 
Assuming an unchanged position of $E_{\text{F}}$ at the substrate interface, the alignment of the Fermi levels in the 
film and substrate results in conduction-band offsets ($\Delta E_{\text{C}}$) of 1.09~eV and 1.02~eV for the films on 
the $n$-type and $p$-type substrates, respectively. Here we have also assumed that the Dirac point is located in the 
center of the \BiSe band gap. Figures~\ref{heterjunct}(a) and~\ref{heterjunct}(b) display the resulting band alignment 
for the \BiSe/Si heterojunction. The upwards (downwards) band bending seen for the $n$-type ($p$-type) substrate is a 
result of the aforementioned pinning of $E_{\text{F}}$ at the surface. Therefore, the energy difference 
$E_{\text{F}}-E_{\text{V}}$ at the substrate side of the interface is the same in the two cases (0.65~eV). For 
simplicity, the band bending across the thin film is assumed to be linear. 

If we instead use the electron-affinity rule~\cite{Anderson1962} to determine the conduction-band offset, we obtain 
$\Delta E_{\text{C}} = 1.14$~eV. The electron affinities $\chi[\text{Si(111)-}(7\times 7)]= 4.16$~eV 
(Ref.~\onlinecite{Hollinger1983}) and $\chi[$\BiSe{}$]\sim5.3$~eV have been used. The latter is an estimate based on the 
value of the work function of \BiSe which is determined from the photoemission experiment. The similar values for 
$\Delta E_{\text{C}}$ obtained using the two approaches indicate that the sketched band alignment is qualitatively 
correct and that the pinned Fermi level position of the substrate is little influenced by the presence of the TI film. 
Also, the magnitude of $\Delta E_{\text{C}}$ at the \BiSe/Si interface is comparable to the band gap of silicon.

Our results show that using Si(111) substrates with very different doping levels does not influence the band bending 
through the \BiSe thin films, as initially expected, due to the pinning of the Fermi level at the interface. On first 
sight, this seems to limit the possibilities of exploiting the bulk position of $E_{\text{F}}$ in the Si substrate to 
modify the electronic configuration of TI thin films and places restrictions on the direction and magnitude of the band 
bending within the TI film as well as the binding-energy position of the Dirac point of the interface state. However, 
this problem might be partly overcome by changing to a hydrogen-terminated Si substrate. At the H-Si(111) surface the 
pinning of $E_{\text{F}}$ is removed~\cite{Karlsson1990,Hunger2002} and the unpinned position is determined by the Fermi 
level in the bulk. Since the conduction-band offset $\Delta E_{\text{C}}$ at the interface is fixed, tuning of the 
chemical potential in the film by Ca or Mg doping~\cite{Hsieh2009e,Chen2010b} is expected to shift the surface Dirac 
point towards lower binding energies. We propose the idea of using a heavily $p$-doped H-Si(111) substrate combined with 
a $p$-doped \BiSe film to obtain the band alignment presented in Fig.~\ref{heterjunct}(c). If the film is sufficiently 
doped, the chemical potential at the surface can be placed below the Dirac point which would result in a hole-like 
surface state and an electron-like interface state as shown in Fig.~\ref{heterjunct}(c). This system could then host 
bound electron-hole pairs, or excitons, and possibly permit the observation of a topological exciton 
condensate~\cite{Seradjeh2009}.

Additionally, we observe that the band bending across the TI film in the \BiSe/Si system, and thus the energy separation 
between the surface and interface states, is larger than for TI films grown on double-layer-graphene-terminated 
6H-SiC(0001) substrates~\cite{He2010a}. Since hybridization effects between electronic states become weaker with 
increasing energy difference, even if there is a finite spatial overlap between the states, the critical film thickness 
above which the surface states on opposite surfaces of a film are decoupled might very well be smaller in the \BiSe/Si 
system as compared to the system described in Ref.~\onlinecite{He2010a}. Systematic studies of the thickness-dependent 
electronic structure in the ultra-thin limit of films grown on different substrates could therefore be of future 
interest. Also beyond the scope of this work are further tests of the topologically protected nature of the states 
interacting with the substrate, for instance the observation of the formation of massive states at the interface as a 
result of breaking the time-reversal symmetry, e.g. in magnetic samples.

In summary, the work presented in this article experimentally confirms the existence of decoupled electronic states 
localized at the interface between a trivial insulator and a topological insulator. Similar to the surface state on 
\BiSe this interface state towards the Si substrate exhibits a Dirac-like linear dispersion. The band bending across the 
TI films, generated by the substrate, separates the Dirac points of the interface and surface states. We anticipate that 
hydrogen termination of the Si(111) substrate will allow the band bending across the film to be manipulated by changing 
the doping level and carrier type of the substrate. Investigations of a TI-TI heterostructure consisting of a 
\BiTe/\BiSe junction shows that the surface state towards vacuum displays a clear \BiTe character while the interface 
state towards the Si substrate remains \BiSe-like. Our data do not provide any evidence for additional states arising 
from the presence of the \BiTe/\BiSe interface. Eventually, our work demonstrates that the combination of 
low-temperature and low-photon-energy ARPES permits direct studies of electronic states at buried interfaces. This opens 
up the possibility for further studies on, for example, interfaces in $p$-$n$ TI junctions and TI-heterojunctions or 
other exotic systems such as the superconductor-TI interface.

\begin{acknowledgments}
This work was made possible through support from the Knut and Alice Wallenberg Foundation and the Swedish Research 
Council.
\end{acknowledgments}

\end{document}